\documentclass[sigconf,nonacm]{acmart}

\AtBeginDocument{%
  }

\copyrightyear{2025}
\acmYear{2025}
\setcopyright{cc}
\acmConference[VINCI '25]{The 18th International Symposium on Visual Information Communication and Interaction }{December 1-3, 2025}{Linz, Austria}
\acmBooktitle{The 18th International Symposium on Visual Information Communication and Interaction (VINCI ’25), December 1-3, 2025, Linz, Austria}
\acmDOI{}
\acmISBN{}

\usepackage{graphicx}
\usepackage{subcaption}
\usepackage{tabularx}  

\begin{document}

\title{GestoBrush: Facilitating Graffiti Artists’ Digital Creation Experiences through Embodied AR Interactions}

\author{Ruiqi Chen}
\affiliation{%
  \institution{The Hong Kong University of Science and Technology (Guangzhou)}
  \city{Guangzhou}
  \country{China}
}
\affiliation{%
  \institution{University of Washington}
  \city{Seattle}
  \state{Washington}
  \country{United States}
}

\email{ruiqich@uw.edu}

\author{Qingyang He}
\affiliation{%
  \institution{Duke Kunshan University}
  \city{Kunshan}
  \country{China}
}
\email{qingyang.he.duke@gmail.com}

\author{Hanxi Bao}
\affiliation{%
  \institution{Duke Kunshan University}
  \city{Kunshan}
  \country{China}
}
\email{hb181@duke.edu}

\author{Jung Choi}
\affiliation{%
  \institution{Duke Kunshan University}
  \city{Kunshan}
  \country{China}
}
\email{jung.choi@dukekunshan.edu.cn	}

\author{Xin Tong}
\authornote{Corresponding author}
\orcid{0000-0002-8037-6301}
\affiliation{%
  \institution{The Hong Kong University of Science and Technology (Guangzhou), Hong Kong University of Science and Technology}
  \city{Guangzhou}
  \country{China}
}
\email{xint@hkust-gz.edu.cn}

\renewcommand{\shortauthors}{Chen et al.}

\begin{abstract}

    Graffiti has long documented the socio-cultural landscapes of urban spaces, yet increasing global regulations have constrained artists’ creative freedom, prompting exploration of digital alternatives. Augmented Reality (AR) offers opportunities to extend graffiti into digital environments while retaining spatial and cultural significance, but prior research has largely centered on audience engagement rather than the embodied creative processes of graffiti artists. To address this, we developed \textit{GestoBrush}, a mobile AR prototype that turns smartphones into virtual spray cans, enabling graffiti creation through embodied gestures. A co-design workshop underscored the role of embodiment—physical engagement with surroundings and body-driven creative processes—in digital workflows. We evaluated GestoBrush with six graffiti artists and findings suggested that embodied AR interactions supporting artists bypass real-world constraints and explore new artistic possibilities, whose AR artworks created enhanced senses of intuitiveness, immersion, and expressiveness. This work highlight how embodied AR tools can bridge the gap between physical graffiti practice and digital expression, suggesting pathways for designing immersive creative systems that respect the cultural ethos of street art while expanding its possibilities in virtual spaces.
    

\end{abstract}


\begin{teaserfigure}
  \includegraphics[width=\textwidth]{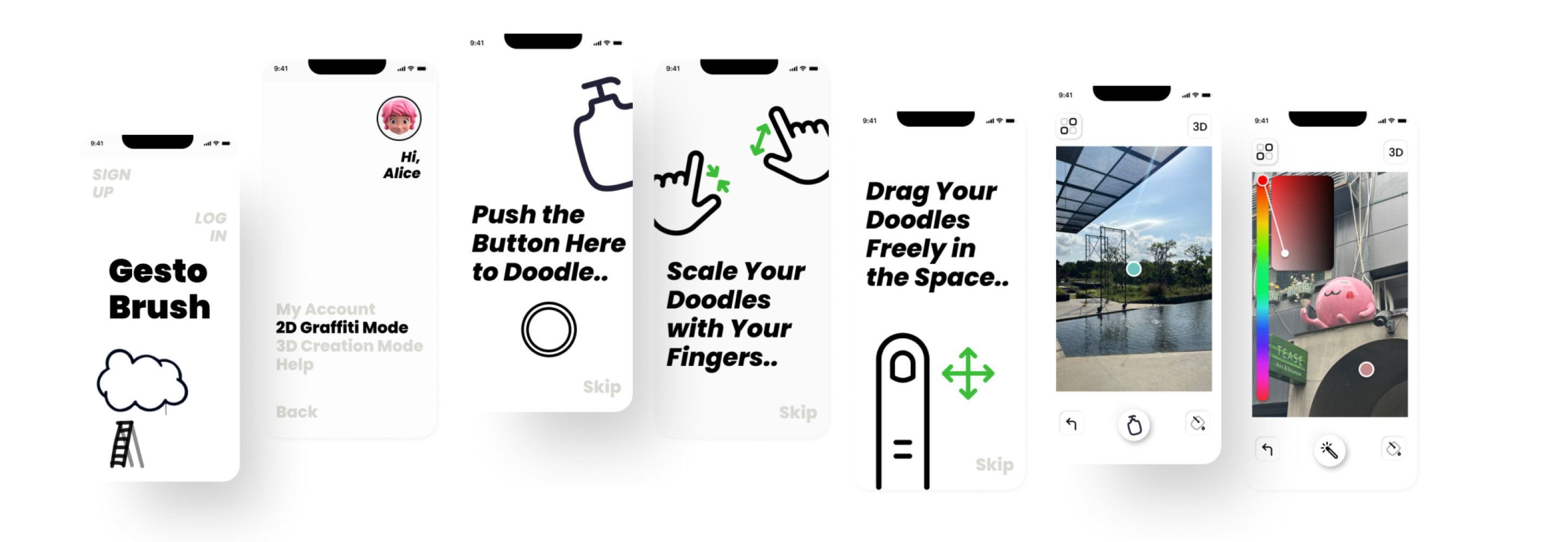}
  \caption{UI design of \textit{GestoBrush}. 
  \textit{Left} section shows the basic UI, including login and mode selection; 
  \textit{Middle} section presents tutorial screens that guide users through embodied AR interactions, such as doodling, scaling, and dragging graffiti; 
  \textit{Right} section illustrates the actual AR graffiti interface, where graffiti is placed and visualized in real-world environments.}
  \Description{The figure shows three groups of UI screens. 
  The leftmost screen displays the basic interface for logging in and selecting graffiti modes. 
  The middle screens show tutorial pages introducing core gestural interactions for graffiti creation, including tapping, pinching, and dragging. 
  The rightmost screens depict the live AR graffiti interface, where graffiti objects are placed into real-world backgrounds and rendered in 3D space.}
  \label{fig:teaser}
\end{teaserfigure}


\maketitle

\section{INTRODUCTION}

Graffiti is a unique form of street art, created by self-driven artists or members of local communities ~\cite{fransberg_embodied_2023, schacter_ugly_2014}. For decades, these “unauthorized” graphical works and stylized letters have documented the evolving socio-cultural landscapes of urban spaces worldwide ~\cite{alonso_urban_nodate, halsey_meanings_nodate, avramidis_reading_nodate}. However, graffiti now faces global restrictions, often perceived as vandalism or a catalyst for crime, which poses concerns for many governments ~\cite{ferrell1997youth, white_graffiti_2001}. 
Since the 2010s, there has been a growing trend toward stricter regulations concerning graffiti in many regions~\cite{moreau_graffiti_2011, landry_stop_2019}, significantly hindering graffiti artists' creative expression and contributing to a gradual decline in graffiti art practices in contemporary society. In HCI research, scholars explored the applications of Extended Reality (XR) technologies in digitalizing and reproducing graffiti artworks~\cite{byeon_roblox_2023, carlonichallenges, di2019disappearance}. Among these, Augmented Reality (AR) stands out for its ability to project digital assets into the physical world, enabling creators to bypass real-world constraints while preserving spatial conditions and cultural contexts ~\cite{geroimenko_augmented_2022, crow2003visible}. Previous AR graffiti studies mainly explored ways in creating interactive AR experiences to enhance audiences’ engagement with the artworks by overlaying digital user-generated content onto real-world surfaces ~\cite{moustafa_augmenting_2023, zuhdi_discourse_nodate}. However, few studies investigated graffiti artists or creators' perspectives to understand their needs in leveraging AR to enhance their digital creative process and the artists' experiences in recreating digital augmented graffiti artworks.


Therefore, we first developed a mobile AR prototype, \textit{GestoBrush}, that takes graffiti artists’ embodied interactions as the drawing actions to enhances their digital creation experiences. We then used \textit{GestoBrush} as a technology probe and conducted an online co-design workshop with 5 graffiti artists to identify the unmet creative needs in their current digital workflows and gain design insights. Findings highlighted that embodiment, a critical element of graffiti creation, was overlooked in the digital creative processes. Embodiment is primarily reflected in two key needs by graffiti artists in their digital creation process: the need of constantly perceiving and engaging with the physical surroundings, and the need of applying their bodily movements and gestural interactions in the digital creation process, mirroring the experience of painting graffiti in physical spaces. 


Therefore, we improved \textit{GestoBrush} for graffiti artists to re-imagine their mobile phones as graffiti spray cans, using their body movements to paint and view AR graffiti in physical spaces through a user interface similar to real-world graffiti interactions via an AR live camera window. Throughout this study, we aim to answer the following two research questions: 

\begin{itemize}
    \item \textbf{RQ1:} What aspects of the current AR graffiti creation process can be improved from the perspective of graffiti artists?
    \item \textbf{RQ2:} How do AR embodied gestural interactions benefit graffiti artists in their AR graffiti creation process?

\end{itemize}

We evaluated \textit{GestoBrush} with 6 local graffiti artists in Tianjin, China. Each participant created one independent AR graffiti work using \textit{GestoBrush} and its creative prompts. Findings suggested that \textbf{GestoBrush}'s embodied AR interactions supported the artists to effectively recreat the graffiti experience in the physical world, allowing artists to use their familiar bodily movements to draw in spatial environments through an intuitive and engaging workflow more. Furthermore, artists noted that integrating graffiti in AR format enabled them to overcome various limitations in reality, such as exploring new perspectives, "painting" at prohibited locations, or addressing regulated themes. Results also showed that the 3D spatial graffiti feature expanded graffiti art from 2D surfaces into 3D space, inspiring new creative possibilities. In conclusion, the contributions of this work are: 1) the system design and implementation of an embodied AR graffiti prototype, \textbf{GestoBrush}; 2) empirical findings from a user study with graffiti artists that evaluated \textbf{GestoBrush}'s potential in fostering embodiment, creativity, and innovation in the digital graffiti creation workflow; and 3) design implications for future research on supporting spatial and gestural embodied interactions in creative AR applications.

\raggedbottom
\section{RELATED WORK}
\subsection{Space, Body, and Graffiti as an Embodied Form of Art}
Due to public space regulations, urban policy, municipal administration, and other factors in modern world ~\cite{anna2011graffiti, mitchell1996city, anderson2012going}, graffiti in physical spaces faces increasing restrictions nowadays ~\cite{moreau_graffiti_2011, white_graffiti_2001}. In HCI, research studies have begun to focus on this issue and explored digital graffiti applications~\cite{byeon_roblox_2023, carlonichallenges, carter2004digital, macdowall2008graffiti}. Yet, as an embodied art form ~\cite{nomeikaite_street_2017, fransberg_embodied_2023}, graffiti is difficult to fully reproduce in digital environments, since its creation relies heavily on artists’ spatial perception and physical movement. It is inherently spatial, both inhabiting and creating environments through practice ~\cite{tuan1977space, simonsen2005bodies}, and rooted in a vernacular interplay between place, sociocultural expectations, and other street visuals such as signage and posters ~\cite{mubi2010wall}. Therefore, the spatial foundations of graffiti art are difficult to replicate within purely digital environments. AR technology has a unique advantage because it can project digital assets onto the physical reality of the real world, preserving its original sociocultural context ~\cite{chmielewska2007framing, valle2010participation}. Furthermore, the creation of graffiti is usually considered a sensuous and embodied experience in which the human body interacts with street artworks and urban space in daily practices, in which creation process, the limbs and torso of the graffiti writer assume a certain shape ~\cite{schacter2016ornament, hannerz2017bodies}, and they produce a particular muscular tension while sketching the enlarged content as a result of performing a gestural routine ~\cite{bowen2010reading, noland_agency_2009}. Some graffiti scholars believe that the creation of graffiti is largely driven by the usual body movements, carrying artistic expression ~\cite{myllyla2018graffiti, nomeikaite_street_2017}. In this process, gestural embodiment gives the artists a sense of agency over their own body in the creation process ~\cite{noland_agency_2009, simonsen2005bodies}, underscoring why this embodied dimension of graffiti is challenging to reproduce digitally.

\subsection{Existing AR Graffiti Practices}
Prior HCI work on XR graffiti has mainly focused on developing XR tools for artists’ digital creation and graffiti digitalization, and building interactive experiences for audiences ~\cite{silva2019ar}. For instance, VR graffiti leverages virtual world features, with applications such as King Spray tailored to graffiti artists ~\cite{pell2017vr, kingspray2023}. In another example, Yu et al. also introduced a prototype combining virtual metaverse technology with graffiti creation, enabling artists to virtually reproduce urban walls or public facilities, create and share works, and express themselves in shared virtual environments ~\cite{zhao2022metaverse}.

Early AR graffiti practices began with Pyksy’s use of Google Tilt Brush in 2017 ~\cite{zuhdi_discourse_nodate, lam2023exploring}, allowing artists to augment digital creations in real-world environments. Recent developments include Simer et al.’s Mobile AR solution for mapping digital graffiti onto 3D surfaces ~\cite{simer_i_nodate}, and Shih et al.’s workflow for creating, geo-tagging, and socially engaging with 3D graffiti, showcasing AR’s evolving role in blending artistic expression with interactive spatial structures ~\cite{shih_grarffiti_2024}. As public understanding and technical access expanded, AR graffiti has emerged as a means of enriching urban visual culture and shared experiences ~\cite{bimber2005spatial, djuric2025enhancing, szabo2020critical}. Biermann’s \textit{‘The Heavy Projects’} initiative exemplifies this potential by using the \textit{Re + Public} AR App to overlay virtual content onto murals, allowing viewers to interact with augmented imagery through their mobile devices in real time ~\cite{BC_theheavyprojects, geroimenko_augmented_2022}. To conclude, existing research primarily focuses on technically supporting artists’ digital creation and providing possibilities for more engaging audience experiences, few research paid attention to graffiti artists' perspectives, e.g., their creative needs and their embodied creation in a digital creation workflow, thus motivating our work. 

\raggedbottom
\section{PRELIMINARY STUDY}
We conducted a preliminary study to better understand graffiti artists’ expectations, needs, and design suggestions for embodied AR graffiti application. The study aimed to identify key challenges in AR graffiti creation (RQ1) and generate actionable insights for system design. We organized a co-design workshop with 5 graffiti artists (2–6 years of experience, varying levels of AR familiarity) and invited them to try a 2D AR graffiti app, followed by semi-structured discussions. In the workshop, participants shared their challenges of using existing AR tools, expectations for embodied 3D AR graffiti creation, and ideas for integrating bodily gestures into their creative workflows. They further elaborated on their graffiti habits, creative processes, and sociocultural influences, while sharing perspectives on AR’s potential to enhance graffiti practice. Insights from this workshop directly informed our prototype design.
\raggedbottom

\textbf{\textit{Bodily Gestures in Graffiti Creation.}}
Body movement plays a crucial role in graffiti practices. According to the participating artists, bodily gestures add a sense of tension and dynamism to the creative process, which enhances their engagement and immersion in graffiti creation. Such movements enable the artists to produce dynamic and fast-paced works, exemplified by techniques like the \textit{station throw-up} (a quick, large-scale graffiti piece done in transit spaces such as train or subway stations, requiring rapid spray movements to finish before being interrupted), which requires rapid and intense bodily actions to complete graffiti pieces within a limited timeframe. Artists also emphasized that their typical tools, such as spray cans and brushes, naturally afford body-tool interaction to create distinctive visual effects. In contrast, they noted that most existing AR graffiti applications constrain users to finger-based drawing on flat screens, eliminating the expressive possibilities and enjoyment provided by bodily gestures. They envisioned that AR technology could enhance the graffiti experience by involving both their hands, upper arms and full-body movements, thereby elevating both the physicality and creativity of digital graffiti creation.

\textbf{\textit{Communication with the Physical Space in Graffiti Creation.} }
Graffiti was also considered as an inherent process of negotiating and communicating with the physical space by the artists. The artists described how creating graffiti often involved exploring and adapting to environmental constraints. Achieving an ideal outcome — or the so-called “perfect draft” — is highly dependent on on-site environmental conditions. Factors such as the surface material of the wall, the available painting space, and the visual harmony between the graffiti and its surrounding environment all play critical roles in shaping the final artwork. Even the same design can appear significantly different depending on its physical context. The artists believed that AR’s spatial expansiveness and interactivity helped overcome these environmental challenges, offering new creative possibilities. They envisioned AR as a tool that allowed graffiti creation to transcend physical limitations, enabling artists to engage more freely with their surroundings. This capability allows for the seamless integration of 2D and 3D graffiti within the physical environment, creating richer and more diverse visual effects.

Building on these insights, we engaged in multiple rounds of brainstorming and iterative design to develop an AR-based graffiti creation application. The following section introduces the final system design, which integrates these key findings into its interaction techniques and spatial features.

\raggedbottom
\section{GESTOBRUSH}

\subsection{Design Features}
\textbf{\textit{Shifting Screen-based AR into Embodied Gestural Interactions:}} 
\textit{GestoBrush} leverages mobile AR technology to translate the traditional graffiti into digital artworks in the environments through dynamic bodily gestures. Informed by our preliminary study findings that emphasizing the role of physical movement in graffiti, we designed a prototype that transforms a handheld mobile device into an embodied tool responsive to full-body gestures. Using Apple’s ARKit and RealityKit, \textit{GestoBrush} supports graffiti creation in both 2D and 3D AR spaces through natural spatial interaction. Unlike conventional AR apps that rely on finger-based inputs, \textit{GestoBrush} enables users to draw using whole-body movements, enhancing engagement with the surrounding environment. This shift from screen-based to gestural control was driven by two design goals: improving spatial perception and immersion by using the phone as a spray can, and reducing reliance on touch interaction to lower barriers for users unfamiliar with AR.
To support this interaction, we designated a fixed point at the edge of the mobile device as the virtual nib of the graffiti tool, simulating the spatial relation found in real-world graffiti practices. By long-pressing a central button on the screen, users can freely move within physical space to draw digital graffiti in AR. \textit{GestoBrush} offers two distinct tool modes: a graffiti spray and a drip mop, allowing users to experiment with different visual styles and effects during the creative process.

\textbf{\textit{Projecting AR into the Physical Space for Seamless Interaction.}}
Our preliminary study also highlighted the critical role of physical space in graffiti creation, emphasizing how artists often negotiate and interact with the environment as part of their creative workflow. In response, \textit{GestoBrush} integrates AR-based environmental recognition to enable users to define and interact with physical surfaces. Users begin by scanning a target wall or surface with their mobile camera, allowing the AR system to register the wall as the canvas. An augmented preview on the screen guides them through the creation process, initially in 2D mode, as shown in Figure ~\ref{fig:teaser} and Figure \ref{fig:example}(B). 
\textit{GestoBrush} then allows users to transition seamlessly from 2D to 3D graffiti creation, moving beyond the limitations of flat surfaces. Once the 2D layer is complete, users can continue drawing into surrounding 3D space, extending their work in real time. This design empowers artists to explore spatial creativity and express ideas more freely, merging 2D and 3D elements within a cohesive AR environment. By doing so, \textit{GestoBrush} expands the expressive range of digital graffiti and supports the creation of dynamic, layered visual effects unconstrained by physical boundaries.


\subsection{Implementation}
\begin{figure}
    \centering
    \includegraphics[width=\linewidth]{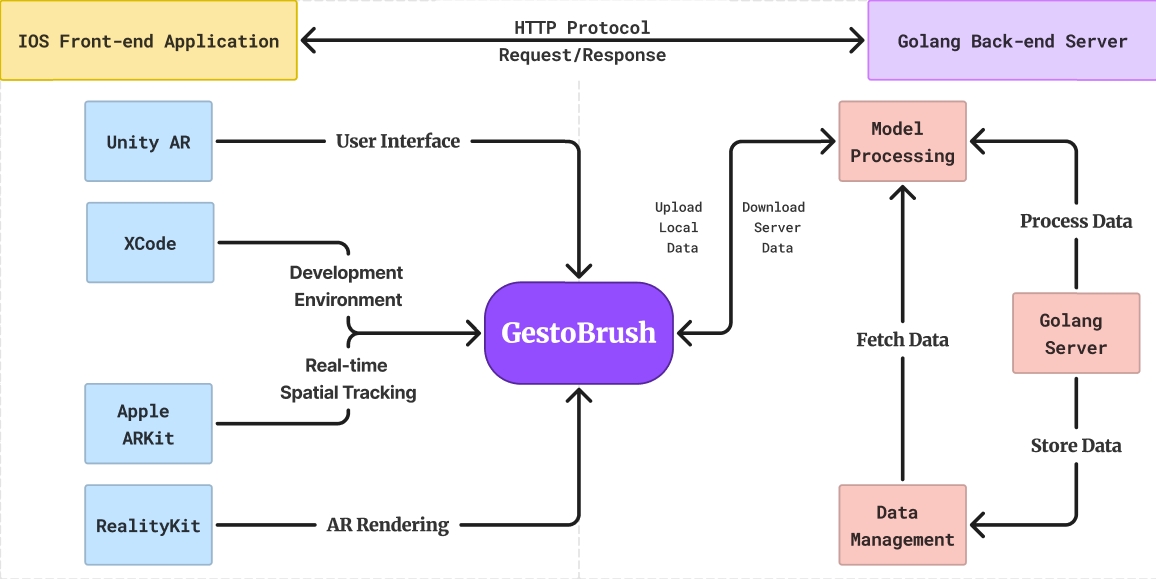}
    \caption{The technical architecture of \textit{GestoBrush} contains a Unity-based iOS front end and a Golang back end to support real-time AR graffiti creation.}
    \label{fig:pipeline}
\end{figure}

The \textit{GestoBrush} system adopts a layered architecture that integrates a Unity-based iOS front-end with a Golang\footnote{\url{https://go.googlesource.com/go}} back-end server to support embodied AR graffiti creation.
On the front end, the application was developed using Unity and Xcode, leveraging Apple’s ARKit\footnote{\url{https://developer.apple.com/augmented-reality/arkit/}} framework to capture real-time spatial and positional data. ARKit is responsible for generating, uploading, and retrieving graffiti models, while RealityKit\footnote{\url{https://developer.apple.com/documentation/realitykit}} enables real-time rendering and interaction within AR space. User-created graffiti can be dynamically generated through a triangular mesh algorithm that processes parameters such as position, size, and color, ensuring accurate placement and expressive customization. The back-end server, implemented in Golang, communicates with the front end via HTTP protocols, which manages the recording, storage, and retrieval of model data, ensuring that graffiti models persist beyond individual sessions and can be reloaded for subsequent use. As illustrated in Figure \ref{fig:pipeline}, this architecture maintains a clear separation of responsibilities; the front end handles user interactions, AR model generation, and rendering, while the back end ensures reliable storage and retrieval. 




\raggedbottom
\section{METHODS}
We conducted a field study with 6 graffiti artists to investigate their experiences and perceptions when using \textit{GestoBrush} for AR graffiti creation. The study consisted of a pre-survey, field sessions of AR graffiti, and two rounds of semi-structured interviews. The pre-survey gathered demographic information such as age, gender, graffiti experiences, and familiarity with AR technology. During the graffiti sessions, participants used \textit{GestoBrush} in real-world environments, while midpoint and final interviews explored their interaction experiences, creative processes, and engagement with AR features.


\subsection{Participants}
We recruited participants by distributing digital posters on Xiaohongshu, a popular social media platform in China widely used by artistic communities. Individuals contacted the researchers directly to enroll in the study. The cohort included participants with varying levels of AR literacy (basic, intermediate, or advanced) and graffiti experiences, ranging from amateur enthusiasts to professional artists (see Table \ref{tab:demographic}). This study was reviewed and approved by Duke Kunshan University's Institutional Review Board. Upon completion of the study, each participant received a ¥100 souvenir as study compensation.

\begin{table}[htbp]
\centering
\small
\caption{Demographic Information of Participants}
\label{tab:demographic}
\begin{tabularx}{\columnwidth}{c c c X X X}
\toprule
\textbf{ID} & \textbf{Gender} & \textbf{Age} & \textbf{AR Tech Literacy} & \textbf{Graffiti Familiarity} & \textbf{Art Experience} \\
\midrule
1 & Male   & 27 & Intermediate  & Advanced   & Professional \\
2 & Male   & 24 & Basic & Advanced   & Professional \\
3 & Female & 23 & Basic & Advanced   & Professional \\
4 & Female & 22 & Advanced    & Intermediate & Professional \\
5 & Female & 25 & Advanced    & Intermediate & Amateur \\
6 & Male   & 25 & Intermediate  & Intermediate & Amateur \\
\bottomrule
\end{tabularx}
\end{table}

\subsection{Procedure}
\begin{figure}
    \centering
    \includegraphics[width=\linewidth]{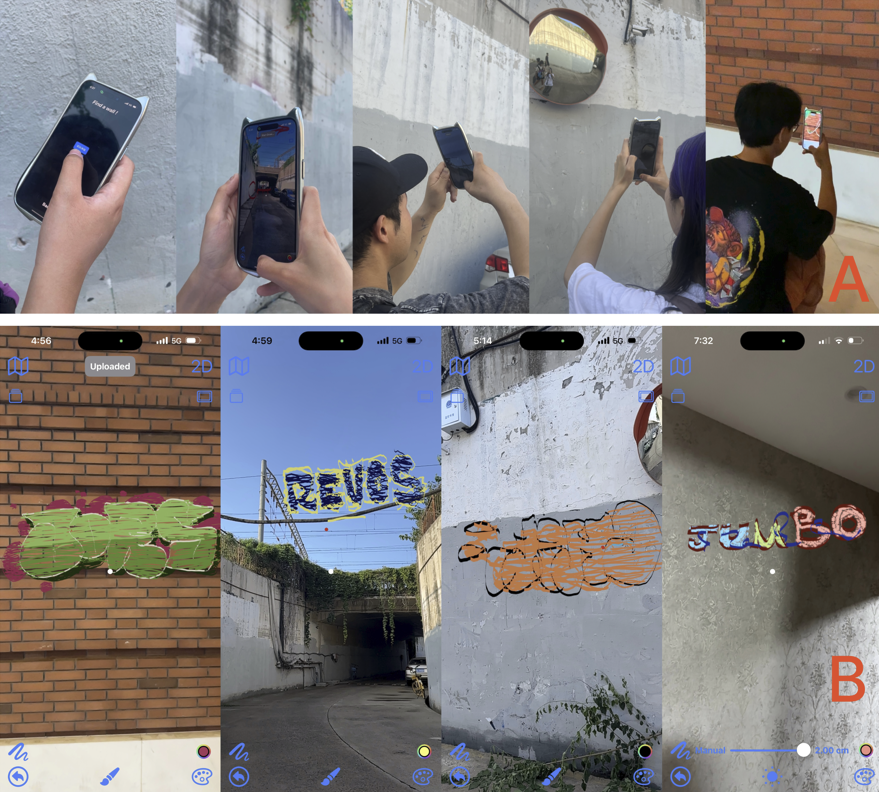}
    \caption{(A): The user flow of \textit{GestoBrush} and multiple views of gestural interactions; (B): The graffiti works created by the artists via \textit{GestoBrush} during the user study.}
    \label{fig:example}
\end{figure}

The study procedure comprised three phases. First, participants attended an online briefing session introducing the study goals, procedures, when researchers obtained their consents of participation. Then, they watched an instructional video about the \textit{GestoBrush} app followed by a live demonstration of a researcher guided them through core features, such as wall scanning and embodied gestural interaction.
The field study was conducted at a locally renowned street known for spontaneous graffiti, chosen to reflect real-world contexts and cultural significance. Participants were encouraged to brainstorm ideas and create an AR graffiti work in 20 minutes (see Figure \ref{fig:example}(A)). For those seeking inspiration, we suggested prompts such as imagining the disappearance of existing graffiti: “What kind of AR graffiti would you leave to commemorate this place and its cultural memory?”



To capture participants’ experiences, we conducted two semi-structured interviews at both the midpoint and the end of the study. Midpoint interviews examined participants’ interactions with \textit{GestoBrush}, the role of AR in shaping their creative practices, and its influence on engagement with the physical environment. The post-study interviews explored their overall usage experiences, comparing 3D AR-based graffiti with traditional graffiti and screen-based AR drawing. 


\subsection{Data Analysis}
We analyzed the interview transcripts using a thematic analysis approach~\cite{clarke2017thematic}. Two researchers independently conducted open coding on transcripts from two randomly selected participants to develop the initial codebook. This codebook was iteratively refined through further transcript analysis and validated using data from two additional participants. Across two coding cycles, the researchers resolved definitional ambiguities and disagreements through discussion until full consensus was reached. The inter-coder reliability (IRR) reached over 80\% between these two researchers.


\raggedbottom
\section{RESULTS}

\subsection{Embodied AR Graffiti Aligns with Real-World Traditional Practices}
Based on insights from our preliminary study, we moved away from traditional 2D AR touchscreen painting and adopted an embodied, gesture-based interaction method for graffiti creation. This new approach was unanimously well-received by all participants. P3 emphasized that this naturally replicates authentic graffiti practice, noting, \textit{``It feels closer to real graffiti because it activates my arms, shoulders, and whole body.''} P2 similarly remarked, \textit{``Using full-body movement reflects the freedom and dynamism of graffiti art.''} Beyond replicating graffiti gestures, participants described embodied AR interaction as heightening immersion compared to conventional 2D screen drawing. As P1 explained, \textit{``This design breaks the limits of flat screens—I constantly interact with the space through body movement, which 2D drawing lacks.''} P6 added that the immersive quality of embodied interaction deepens engagement with graffiti’s physical language: \textit{``It lets me explore different angles and spaces, encouraging artistic engagement with the environment.''}

However, while embodied interaction in \textit{GestoBrush} was generally well received, participants also noted challenges and a learning curve. Several artists mentioned that adapting to 3D spatial control with a mobile device required practice. As P2 commented, \textit{``At first, I had trouble aligning my movements with the spray's position in 3D space.''} P4 similarly observed that small body motions could cause unintended strokes. P5 also raised concerns about spatial constraints, \textit{``AR graffiti needs physical space, which isn’t always available—especially in tight or crowded places.''}
At the same time, participants emphasized that embodied AR interaction enriched the expressiveness of their work, enabling creative output and emotional communication through body language. P4 said, \textit{``Through body movements, I convey emotions and ideas—and the audience can sense them too.''} P5 echoed this view, saying that \textit{``Viewers don’t just see the final piece—they witness the creation process and better understand the emotions behind it.''}
In summary, while embodied AR interactions align with real-world graffiti practices and foster a more immersive, expressive experience, they also introduce challenges to the artists, such as learning curves, control precision, physical fatigue, and spatial constraints. These findings revealed important design implications for future AR creative tools, where balancing immersion with usability remains a critical consideration.

\subsection{Overcoming Physical Spatial Constraints with Embodied AR Graffiti}
In addition to enabling embodied interaction, participants highly valued \textit{GestoBrush}’s capacity to transcend the physical constraints of traditional graffiti. They emphasized the freedom to create beyond walls, fixed locations, or environmental conditions. P4 reflected: \textit{``In traditional graffiti, I had to adapt to the environment—wall size, lighting, shadows—which limited my inspiration. But in \textit{GestoBrush}'s AR space, I can create freely and boldly, as I envision it.''}
Participants also described how embodied AR graffiti reshaped their understanding of graffiti spaces, opening possibilities for unconventional locations and orientations. As P3 noted, \textit{``This embodied AR approach lets us paint anywhere—not just on walls, but even in mid-air or the sky.''} P1 further elaborated: \textit{``It broadens our work beyond physical spaces, integrating the socio-cultural context. We're no longer confined by specific sites or themes in AR digital space.''}

However, while this spatial freedom was broadly appreciated, participants also reflected on challenges of creating in an unconstrained virtual space. Several worried about losing the strong connection to real-world locations, which have traditionally shaped the meaning and cultural significance of graffiti. As P4 explained: \textit{``When there’s no physical boundary, the graffiti sometimes feels like it's floating—location usually gives the piece its meaning.''}
Others noted that the absence of physical surfaces sometimes made it harder to control positioning or alignment, leading to moments of disorientation. P2 remarked: \textit{``Without a real wall as reference, it was hard at times to judge scale or placement—especially in wide, open areas.''}
Participants also pointed out technical limitations. Outdoor lighting or a lack of textured surfaces occasionally reduced AR tracking stability. As P5 shared: \textit{``In bright outdoor spaces or areas lacking visual features, the AR graffiti sometimes drifted or failed to stay anchored.''}

Despite these challenges, participants generally felt that creating graffiti beyond physical walls encouraged new ways of thinking about space and art inspirations. P5 reflected: \textit{``It encourages artists to treat the environment as part of their work with a focus on spatial forms, objects, and lighting, not only using space as a passive canvas.''} P6 further suggested that digital AR graffiti can become an important direction for preserving graffiti culture amidst growing restrictions on physical graffiti: \textit{``As graffiti artists, we’ve always relied on physical walls. This AR app offers a new way to preserve graffiti culture, especially when more places restrict wall-based graffiti.''}

\subsection{Innovation and Inspiration through 3D Spatial Graffiti in Embodied AR}
Although \textit{GestoBrush} introduced transformative changes to traditional graffiti—such as embodied spatial AR interaction and the ability to break free from physical wall constraints—all participants agreed that its most significant contribution lies in pioneering 3D spatial graffiti. This innovation redefines the boundaries of graffiti as an art form and establishes an entirely new mode of digital expression. As P2 reflected, \textit{``I never imagined creating graffiti in a 3D digital space. It completely changed how I think about both traditional and AR-based graffiti.''}

Participants highlighted how this method enhances the layering and spatial depth of graffiti, making it more dynamic and visually engaging. As P1 mentioned: \textit{``I can draw from any position or angle in 3D space—no blind spots. It feels refreshingly new!''} They also emphasized the unique experience of viewing their work from multiple perspectives, which becomes part of the design process itself. P3 explained: \textit{``This new audience perspective helps me understand and present my work from different angles. I can incorporate these views into the design.''}
Beyond boosting interactivity and immersion, participants described how 3D graffiti inspired them throughout the creative process, opening new avenues for artistic expression. Several also stressed the sense of freedom this mode affords, as it vastly expands the creative space and fully unleashes artistic potential. As P6 explained: \textit{``It breaks free from 2D limitations and truly brings graffiti into 3D space. It gives artists much more freedom, letting them adapt and shape their work based on the environment—which aligns perfectly with graffiti’s nature as an in-the-environment art form.''}

Some participants observed that, unlike traditional flat-surface painting with brushes, \textit{GestoBrush} offers a fresh and dynamic creative experience by transforming the smartphone into a graffiti brush and 3D space into the canvas. As P2 explained: \textit{``The smartphone, a portable and familiar everyday object, has now become a tool for digital graffiti. This experience is inherently paradoxical. \textit{GestoBrush}'s interaction design blends the familiarity of smartphone use with the physical movements of traditional graffiti. As graffiti artists, this feels familiar and novel.''}
P5 highlighted that this innovative interaction design not only sustains users’ interest in digital graffiti but also motivates them to experiment and create. He further suggested that its appeal extends beyond graffiti practitioners to include individuals with no prior experience, thereby broadening the art form’s audience and possibilities: \textit{``Using a familiar tool like a smartphone as a brush—and recording and sharing their creations—could spark curiosity and inspire people to create, unlocking their creative potential.''} Taken together, these reflections underscore how \textit{GestoBrush} not only reimagines graffiti through embodied AR but also expands its accessibility and expressive potential, establishing 3D spatial graffiti as a distinctive and transformative artistic practice.

\raggedbottom
\section{DISCUSSION}
Our study found that \textit{GestoBrush} successfully replicated real-world graffiti practices through embodied AR interactions, allowing artists to create intuitively while overcoming physical constraints and extending graffiti into 3D spatial forms. Despite these contributions, the study was limited by a small sample size and lack of controlled settings, leaving improvement space for future work.

\subsection{Design Implications}

\subsubsection{Adding embodiment gestural interactions in the AR space}
Our findings revealed that \textit{GestoBrush}’s gestural interaction design significantly enhanced artists’ sense of embodiment in digital creation workflows. This builds on prior research aimed at simulating graffiti artists’ real-world environments and practices in digital settings \cite{byeon_roblox_2023, di2019disappearance,zuhdi_discourse_nodate}, enabling a closer replication of the physical act of graffiti creation. In particular, our study shows that engaging full-body movements and dynamic postures allowed artists to interact more continuously with the creation space and immerse themselves more fully in the painting process.

These findings align with prior work highlighting the role of bodily perception and schema in shaping interaction design, beyond purely graphical or traditional physical interfaces ~\cite{dourish2001action}. We encourage future research to explore how creative AR systems can further leverage the interplay of body movement and digital interaction, and how gestures can be made more perceptible and meaningful within AR environments ~\cite{gavgiotaki2023gesture}. At the same time, participants reported a learning curve in adapting to the embodied model, particularly in accurately controlling the position and scale of graffiti strokes in 3D space. The physical demand occasionally led to fatigue or discomfort, interrupting their creative flow.

For example, future embodied AR prototypes could integrate advances in tangible and social computing to heighten users’ awareness of their physical bodies and surroundings. Such integration aligns with the vision of AR as a medium for seamlessly blending digital content with the physical world ~\cite{chen2012kinetre, kim2016sketchingwithhands}. Further exploration could also focus on multi-sensory feedback, e.g., visual, auditory, or haptic cues, to enhance immersion by responding directly to users’ movements in AR spaces ~\cite{krueger1985videoplace}. Another key insight from our findings is that replicating real-world gestures can supports intuitive control and enables richer emotional and creative expression. Thus, we suggest that future creative AR systems consider how embodied gestures, especially when situated in specific physical places, can foster deeper immersion, emotional engagement, and contextual awareness ~\cite{benford2006can, de2017pokemon}.

\subsubsection{Breaking the real-world regulations via physical augmentation}
Our findings show that integrating graffiti with AR technology enabled artists to overcome many spatial and regulatory constraints in traditional graffiti. Participants particularly valued the ability to create graffiti from previously inaccessible perspectives and locations, expanding their creative freedom beyond physical walls. In this sense, AR-based digital graffiti extends the boundaries of the art form, allowing artists to incorporate physical surfaces, surrounding space and digital elements into their creative environments.

This insight aligns with prior research showing how virtual environments can emulate or customize graffiti conditions, such as color, dripping effects, and lighting, to create more engaging digital experiences ~\cite{zuhdi_discourse_nodate, xu_pre-patterns_2011}. Similarly, \textit{GestoBrush} enabled artists to transcend spatial limits, challenge norms, and create without being tied to specific sites or themes. This resonates with literature on AR-supported activism, which highlights AR’s capacity to bypass physical constraints and deliver immersive, multilayered narratives that surface hidden histories and perspectives ~\cite{silva_understanding_2022, moustafa_augmenting_2023}. At the same time, our study revealed new challenges. Several artists described disorientation and difficulty maintaining contextual relevance without real-world surfaces. The lack of physical boundaries complicated spatial judgment and reduced the cultural meaning often tied to location-based graffiti. Technical limitations—such as unstable AR tracking under certain conditions or insufficient space for full-body gestures—also occasionally disrupted the experience.

These findings suggest that while AR graffiti systems hold strong potential for supporting self-expression, free speech, and creative activism in virtual spaces, designers must also consider ways to help artists preserve spatial coherence and contextual grounding. For example, participants noted opportunities to naturally integrate environmental features—such as spatial structures, nearby objects, or lighting conditions—into their AR graffiti, fostering a more seamless relationship between digital content and the physical world.


Therefore, future research could explore ways to reshape the user-place relationship in AR graffiti, through expanding creative freedom and enhancing the perception of spatial affordances and strengthening the connection between users and their physical environment ~\cite{azuma2017making, herrera2018building}. Such efforts can help artists navigate the tension between virtual freedom and contextual relevance, fostering richer and more situated creative experiences in embodied AR environments.

\subsubsection{Transforming the conventional 2D graffiti art into 3D digital experiences}
Beyond replicating real-world graffiti creation practices, \textit{GestoBrush} also introduces new possibilities for re-imagining graffiti as a spatial artistic experience, transforming its conventional form from a 2D-based tangible practice into dynamic and embodied 3D digital creations. Our findings suggest that the embodied AR creation process enhances the layering, spatial dimensionality, and cultural contextualization of graffiti artworks, making them more immersive and visually engaging for both artists and audiences.

We encourage future research to further explore how embodied AR technology can be designed as an immersive medium for interacting with spatial structures, thereby shaping the potential of AR graffiti and its contribution to a shared urban visual language and culture ~\cite{fisher2021augmented, sanaeipoor2020smart}. In particular, 3D spatial graffiti creation can serve as an engaging and interactive approach to fostering collective creative experiences in public spaces. For instance, in the context of co-creative AR prototypes, the Dream Garden system developed by Petrov et al. ~\cite{petrov2023dream} enables users to collaboratively place virtual 3D flowers within the physical environment, facilitating social communication and interaction in public settings. Similar practices could be incorporated into the gestural embodied interactions of \textit{GestoBrush} to support more interactive, collocated experiences among urban users, leveraging body language and movement as a means of connection and communication. Moreover, future research could investigate how embodied 3D AR experiences might be integrated with location-based social networks (LBSNs) ~\cite{mcgookin2014studying, herskovitz2022xspace}, further expanding the potential of AR graffiti to cultivate a broader, more vibrant, and globally connected creative community.

\subsection{Limitations}
While this study highlights the potential of embodied AR interactions, several limitations remain. The small sample size of six participants based in China, may not reflect the full diversity of global graffiti communities. Cultural and geographic differences can influence how artists engage with AR tools, warranting broader future studies. Additionally, the study was conducted in controlled settings with guided prompts, which may not fully capture the spontaneity and improvisation central to traditional graffiti. Field research in more natural environments is needed to assess real-world applicability. 
\raggedbottom

\section{CONCLUSION}
In thie study, we explored how embodied graffiti AR technology can address challenges in graffiti artists’ digital workflows and developed a prototype named \textit{GestoBrush}. Findings showed that \textit{GestoBrush} effectively recreated the physical graffiti-making experiences, enabling artists to use familiar embodied gestures, engage with their surroundings, and create digital graffiti in restricted locations or from new perspectives. These findings offer valuable design insights for future embodied AR prototypes, emphasizing intuitive interaction, spatial engagement, and creative flexibility for AR graffiti creators.


\raggedbottom

\bibliographystyle{ACM-Reference-Format}
\bibliography{manuscript} 


\end{document}